# Intervening BLR Clouds' Effects on Optical/UV Spectrum

Ye Wang[1], Gary J. Ferland[1], Chen Hu[2], Jian-Min Wang[2], Pu Du[2]

[1]Department of Physics and Astronomy, 177 Chemistry/Physics Building, University of Kentucky, Lexington, KY 40506, USA

[2]Key Laboratory for Particle Astrophysics, Institute of High Energy Physics, Chinese Academy of Sciences, 19B Yuquan Road, Beijing 100049, China

Abstract

Recent x-ray observations of Mrk 766 suggest that broad emission line region clouds cross our line of sight and produce variable x-ray absorption. Here we investigate what optical/ultraviolet spectroscopic features would be produced by such "Intervening BLR Clouds" (IBC) crossing our line of sight to the accretion disk, the source of the optical/UV continuum. Although the emission spectrum produced by intervening clouds is identical to the standard BLR model, they may produce absorption features on the optical or UV continuum. Single clouds will have little effect on the optical/UV spectrum because BLR clouds are likely to be much smaller than the accretion disk. This is unlike the X-ray case, where the radiation source is considerably smaller. However, an ensemble of intervening clouds will produce spectroscopic features in the FUV including a strong depression between the Lyman limit and Ly$\alpha$. The amount of the depression will indicate the line-of-sight covering factor of clouds, an unknown quantity that is important for the ionization of the intergalactic medium and the energy budget of AGN. Comparison with observations suggests that the SED of Mrk 766 may be affected by intervening BLR clouds and IBC may exist in most of AGNs.

INTRODUCTION

The nature of the Broad Line Region (BLR) of Active Galactic Nuclei (AGN) is an area of vigorous debate (see the summary by Wang et al. 2011, Wang et al. 2012, Osterbrock & Mathews 1986, Korista 1999, Ferland 2004). The BLR is used to measure black hole masses and may be the avenue coupling feedback between the central black hole and the surrounding galaxy. The picture that has emerged is one where a large number of clouds orbit the central black hole (Arav et al. 1997, 1998). Some recent work on the structure of AGN suggests that the clouds, which emit the broad lines in the UV and optical, constitutes a flattened, self-shielding and physically thick system (Gaskell, Goosmann & Klimek 2008), as Figure 1 shows. Indeed, it may provide the coupling between the outer regions of the accretion disk and the inner parts of the obscuring dusty torus (Wang et al. 2010). The accretion disk, BLR, and torus may be a continuous geometry with matter flowing from the torus through the BLR, onto the disk and eventually the black hole (Hu et al. 2008a, 2008b). The geometry of individual BLR clouds, where the strong emission lines form, is also uncertain. Some models that invoke winds off the accretion disk (Wang et al. 2012), or a fine "mist" of small clouds (Arav et al. 1998), have been proposed.

Some recent work (Risaliti et al. 2005, 2007, 2009, Elvis et al. 2004, Puccetti et al. 2007, Bianchi et al. 2009, Bianchi, Maiolino & Risaliti 2012) suggests that BLR clouds may produce X-ray absorption, which is very common in AGNs. In this work, X-ray absorbers are described as clouds with linear dimension of the order of $10^{13}$-$10^{14}$ cm, velocities in excess of $10^3$ km s$^{-1}$ and densities $n \approx 10^{10}$ - $10^{11}$ cm$^{-3}$. The distance from the X-ray source to the absorbers would be of the order of thousands of gravitational radii, assuming that the clouds orbit the black hole with Keplerian velocities. The similarity of these physical parameters to those of the BLR clouds suggests that they are the same.

Observations of Mrk 766 (Risaliti et al. 2011, hereafter R11) and NGC 4395 (Nardini & Risaliti 2011) suggest that variations in the observed properties of the X-ray absorption are due to the motion of X-ray absorbers across the line of sight (LOS) between the observer and the X-ray source. R11 also argued that the absorbing clouds have denser neutral core, which absorbs the X-rays, and an extensive "comet tail" of high ionization gas.

The BLR global covering factor should be, on average, 40% (Dunn et al. 2007, Gaskell, Klimek & Nazarova 2007, Gaskell, Goosmann & Klimek 2008). Because of the large vertical thickness of the torus, we usually view most type -1 AGNs within ~ 45 degrees of the polar axis (Smith et al. 2004). Considering these specifics of the BLR geometry, we may view the central regions of most AGNs through the BLR. The purpose of this paper is to examine the observational consequences of such geometry.

There have been a large number of studies of the emission properties of BLR clouds. The most sophisticated are the "LOC" models that allow properties of clouds to change within the BLR (Baldwin et al., 1995). These models largely reproduce observed emission properties. This spectrum is not changed by the geometry proposed in the Risaliti et al. work. Rather, the novel aspect of this work is to consider the absorption produced by emitting clouds. Some clouds lie along the line of sight to the central accretion disk and black hole and produce certain characteristic absorption features on the accretion disk continuum. We show what effects clouds have on the optical and ultraviolet continuum, and identify a broad feature below Lα that appears to be present. This has implications for the true SED of the accretion disk.

2 Parameters and geometry of clouds

2.1 Parameters of clouds and continuum source

To know what effect the clouds will have on the observed spectrum we need to compare the size scales of a cloud with the size of the continuum source. For instance, if the cloud is larger than the continuum source, which is likely to be true at X-ray energies, a single cloud can nearly fully block our view of the X-ray source. However, if the cloud is much smaller than the continuum source, as we show seems likely for the optical or ultraviolet, a single cloud may have little effect upon the spectrum and we would mainly observe the effects of an ensemble of clouds.

We assume that the parameters derived by R11 and mentioned in the introduction are reasonable enough for a typical BLR cloud. Therefore we assume a cloud density of $n$(H) = $10^{10}$ cm$^{-3}$ and that the cloud column is $N$(H) = $10^{23}$ cm$^{-2}$. The corresponding linear scale of the cloud is $10^{13}$ cm.

R11 used XMM- Newton observations of Mrk 766 to derive the size of the X-ray source. They did this by combining the transverse velocity of the clouds with their occultation times derived from observed variability. They estimate that the X-ray source has a size $r_{xr}$ between $20R_g$ and $100 R_g$, based on the highest physically acceptable cloud velocity. They further argue that the source is likely to have a size closer to the lower end of the range. The range corresponds to between $r_{xr}$ ~$10^{13}$ cm and a maximum of $r_{xr}$ ~$10^{14}$ cm for Mrk 766.

This physical size of the X-ray source is close to the cloud length scale derived above (R11). As a result a single cloud can easily block the X-ray source and strongly absorb the X-ray spectrum, letting very little light pass to the observer if the cloud column density is large enough. This is the essence of the R11 model.

In this paper we ask what these intervening clouds will do to the UV/optical spectrum. The radiation field in this spectral region is predominantly emitted by the accretion disk surrounding the central black hole. For AGN Mrk 766 the parameters are black hole mass $M_6$ (1 $M_6$=$10^6$ $M_\odot$) =$1.76^{+1.56}_{-1.40}$(Bentz et al. 2009) and a luminosity in the range $L$=$10^{44.0}$ to $L$=$10^{44.4}$ erg/s (Vasudevan et al 2010; we assume $L$=$10^{44.2}$ erg/s). These correspond to an accretion rate of $\dot{M}$=0.03 $M_\odot$ year$^{-1}$ (Equation 3.7 in Peterson 1997). The corresponding Eddington ratio is 0.7.

We next estimate the size of the UV/optical continuum source with these parameters. There are different analytical forms of the temperature – radius relation for disks around black holes. Bregman (1990) gives an expression for the temperature of gas near the last stable orbit. This will be at far smaller radii than the parts of the disk which emit in the FUV and optical. Wang & Zhou (1999) give expressions for super-Eddington accretion. As shown above, the Eddington ratio is less than unity, although there are substantial uncertainties and the ratio may be larger than one. For simplicity, we assume a sub-Eddington thin disk that satisfies the standard α-disk model (Shakura & Sunyaev 1973) and assume α=0.1. For parts of the disk dominated by gas pressure and free-free absorption, the relation between disk temperature and radius (equation A11 of Shields & Wheeler 1978) is approximately given by

$$T = (4.3\times10^6 K)(\alpha^{-6/23}M_8^{-10/23}\dot{M}^{8/23})R_g^{-15/23} \qquad (1)$$

where $R_g$ is the Schwarzschild radius.

We assume that the accretion disk radiates as a blackbody. Since the temperature is a function of the radius of the disk, we can consider that the luminosity of light at a particular frequency is also a function of the accretion disk's radius. To describe this function, we introduce the differential luminosity. This presents the intensity at a particular frequency on one side of the accretion disk ring, whose center is the black hole and width is a unit length. Assume $r$ is radius, $v$ is frequency, $\Delta r$ is unit length, and $B_v(T) = (2hv^3/c^2)(\exp(hv/kT) - 1)^{-1}$ is Planck's Law. The differential luminosity is then

$$L_v(r) = 2\pi r \, \Delta r B_v(T) \qquad (2)$$

If $T_0$ is the temperature at some radius $r_0$, then the ratio $\eta_v(r) = L_v(r)/L_v(r_0) = (r/r_0)(B_v(T)/B_v(T_0))$ is the relative differential luminosity and is plotted in Figure 2. This shows that light at a frequency rises to a peak at some radius and then vanishes rapidly. For simplicity, we can assume that the source of light at that frequency is the accretion disk at the radius of maximum emission. With this assumption, the linear scale of the UV/optical continuum source is $10^{13}$~$10^{15}$ cm, several to hundreds of times large than clouds' scale. This has important consequences for what we observe at UV/optical wavelengths.

2.2 Global and line of sight covering factors

The global covering factor is the fraction of sky as seen from the central engine that clouds cover. This can, for instance, be obtained from the equivalent width of the lines (Korista, Baldwin & Ferland 1997). A typical global covering factor is 40% (Dunn et al. 2007, Gaskell, Klimek & Nazarova 2007, Gaskell, Goosmann & Klimek 2008).

However, when we observe a galaxy, the absorption effects of clouds upon the spectrum of the central engine is determined by the line of sight (LOS) covering factor, the faction of the continuum source obscured by clouds. If the geometry is spherically symmetric then the global and LOS covering factors will be the same. However the geometry of an AGN is thought to have cylindrical symmetry, so a particular LOS may or may not intersect clouds. The LOS covering factor may be larger or smaller than the global covering factor, depending on where clouds lie with respect to our line of sight. The R11 results do suggest that clouds occur on the LOS for the objects they studied. In the remainder of this paper, we will only consider the LOS covering factor.

2.3 Geometry of clouds

Here we will show the spectroscopic differences between two models: the classical standard model that assumes that no clouds lie on the LOS and the R11 model in which some clouds lie along the LOS. We will refer to the second as the Intervening BLR cloud model (IBC model) in the remainder of this paper. Differences in the predicted spectra could be used to determine the LOS covering factor.

The standard model (Figure 3) presents the geometry most often assumed in the literature. We receive light directly from the central continuum source and have an unobstructed view of it. The clouds, which are energized by the central engine, do not lie along the sight line between observers and the central engine.

This model was motivated by the fact that the column density in a typical BLR cloud is large enough for it to be quite optically thick in the $H^0$ Lyman continuum. The standard model was originally suggested by the fact that we do not observe the strong Lyman limit absorption that would be produced by clouds between the observer and the central engine. We show below that realistic BLR clouds do not produce a feature at the Lyman limit due to the presence of other opacity sources.

The IBC model is shown in Figure 4. As in the standard model, the clouds reprocess the radiation field emitted by the central engine, but some occur along our sight line to central regions. The cloud emission properties are unchanged, but intervening clouds can absorb parts of the continuum produced by the central regions. Intervening clouds can change the observed spectrum, especially the SED of the central engine.

Below we quantify differences in observed properties of these two models as a way to test what actually occurs.

3 Photoionization Calculations

3.1 Calculations

It is not the goal of this paper to create full models of the BLR such as the locally optimally emitting clouds (LOCs) picture of Baldwin et al. (1995). Rather we shall use typical values for cloud parameters, as given, for instance, by Peterson (1997) or Korista, Baldwin & Ferland (1997). Since the R11 geometry will not alter the emission-line spectrum, we concentrate on the absorption produced by the clouds. We assume that the clouds have a hydrogen number density of $n$(H) = $10^{10}$ cm$^{-3}$. R11 infer two components to a cloud, a core and halo, from their X-ray analysis. Accordingly, we consider clouds with column densities ranging between $N$(H) = $10^{21}$ and $10^{23}$ cm$^{-2}$. Similarly we consider a range in distance between the central black hole and the cloud between $r$ = $10^{15}$ and $10^{17}$ cm. The effect of varying both density and radius is to cover parts of the LOC plane described in Korista et al. (1997). We vary the column density because we are mainly interested in determining the UV/optical absorption properties. Korista et al. (1997) considers the effects of various column densities in detail. As we show below the range of $N$(H) does change the observed absorption, while previous work has shown that it has only a weak effect on the emission spectrum (Korista et al. 1997).

We simulate the spectrum with version 10.0 of the plasma code Cloudy, last described by Ferland et al. (1998). The properties of such photoionization models are described by Osterbrock & Ferland 2006, Ferland 1999, and Ferland 2003. These properties are not changed by the assumptions introduced in this paper. We adopt the observed luminosity of Mrk 766 of $10^{44.2}$ erg/s given above and use the Mathews & Ferland (1987) SED. Solar abundances are assumed. We will show and discuss the results in following sections.

The predicted X-ray spectra are shown in the right side of Figures 5, 6 and 7. These correspond to clouds at distances from the black hole of $10^{15}$ cm, $10^{16}$ cm, and $10^{17}$ cm respectively. The X-ray spectrum is nearly unchanged by clouds at the smaller radius, expect for a few absorption lines produced at high column densities. As the radius increases the ionization of the gas goes down and X-rays absorption increases. These figures also show the well-known effect that lower-energy photons, with longer wavelength, are more easily absorbed due to the larger gas opacity.

The UV/ optical spectra are shown in the left sides of the figures. Clouds close to the continuum source have little effect due to their high ionization and low opacity. Emission and absorption lines begin to appear as the radius increases and the ionization decreases. These features also become stronger when the cloud column density is increased. The UV/optical absorption spectra, the new result of this paper, are described next.

These results suggest that intervening BLR clouds could affect the X-ray and UV/optical spectra of AGN in different ways. We next discuss a single cloud's effect on the spectrum; then consider the BLR cloud as the X-ray absorber; and finally give the UV/optical spectra of an ensemble of clouds that lie along our sight line. We assume all clouds have the canonical BLR parameters for simplicity (Korista et al. 1997). We use this representative cloud's spectrum and combine this with the covering factor to evaluate the net spectrum.

3.2 A single cloud's effect on the spectrum

Clouds will have very different effects on the X-ray and UV/ optical spectra because of the significant differences between the linear scales of the clouds, the X-ray source, and the UV/optical source. The X-ray source is smaller than ~ $10^{13}$ cm (R11), so that a single cloud can obscure a considerable part of the X-ray source, as R11 point out. A single high column density, low ionization, cloud will produce strong absorption, leading to remarkable X-ray time variability when it crosses our sight line.

The simple α-disk computed above (Figure 2) suggests that the UV/optical source has a linear scale of ~$10^{13}$-$10^{16}$ cm, which is several to hundreds times larger than the cloud. Because of this size difference, a single cloud passing in front of the accretion disk will have no observable affect since it would only cover 10% to 0.0001% of the disk's visible surface. However, as Arav et al. (1997; 1998) point out, there is likely to be an ensemble consisting of a large number of clouds. Although individual clouds may come and go along our sight line, it is most likely that there is no net change in the total ensemble of clouds along this sight line. The ensemble of clouds will not produce time variations in UV/optical band. However, if the ensemble has a global covering factor of ~40%(Dunn et al. 2007, Gaskell, Klimek & Nazarova 2007, Gaskell, Goosmann & Klimek 2008) and if LOS covering factor is similar to this value, the clouds will change the observed SED. We show their affect upon the observed spectrum below.

3.3 The BLR cloud as the X-ray absorber

Which BLR clouds are responsible for the X-ray absorption? The absorption depends strongly on the state of ionization. If the cloud is highly ionized, there is very little X-ray absorption, as shown in Figure 5 and 6. These clouds are near the black hole and have higher ionization than the more distant clouds in shown Figure 7. The high-ionization clouds have little spectroscopic effect except to Compton scatter a small fraction of the incident radiation field. But the more distant low-ionization clouds strongly absorb at X-ray wavelengths. We also show this in Figure 8 by comparing the transmitted spectrum, the light we observe looking through a cloud, with its original intensity. Here we use the X-ray deduced covering factor of 40% (R11 estimate that this is between 20%-80%). The graph shows that X-ray absorption becomes obvious for clouds with column densities of $\sim 10^{22}$ cm$^{-2}$-$10^{23}$ cm$^{-2}$ and for clouds that are farther from the black hole. This means that lower ionization clouds have stronger X-ray absorption, and will be more distant since these have the larger opacity. This provides an explanation for the R11 conclusion that the obscuring clouds have neutral cores with column densities of a few $10^{23}$ cm$^{-2}$ and is surrounded by a highly ionized halo.

3.4 The UV/optical spectra of clouds along our line of sight

Next consider the effects of the same clouds on the UV/optical spectrum. These clouds generate strong emission lines in this band, which is a defining feature of BLR clouds. But intervening BLR clouds will also produce absorption, as described here. We focus on the spectral region between 800Å and 3000Å, where intervening clouds have the most obvious effects.

We first consider the simplest case where clouds fully cover the continuum source. We consider a typical BLR cloud with $n$(H)=$10^{10}$ cm$^{-3}$ and $N$(H)=$10^{23}$ cm$^{-2}$ (Korista et al,. 1997) as representative. The resulting spectrum is shown in Figure 9(b). The incident spectrum, light directly received from the continuum source, is shown as the red line, and the transmitted spectrum, which is the light that passes through the clouds, is the black line. Many absorption features are visible. The lower panel in Figure 9 shows the continuous optical depths integrated across the cloud. This illustrates the atomic processes that produce the features shown in Figure 9(b). In 9(c), the total optical depth is shown in red. The scattering optical depth, dominated by Lyα, is shown in blue. The cloud is quite optically thick to Lyα scattering, which removes photons in the neighborhood of the line. There are three continuous absorption features shown in black. There are due to photoionization of ground state H$^0$ (912Å), O I$^*$ (~1060Å this is the first excited $^1$D term of O I, Ralchenko et al. 2011), and a weak feature due to ground state photoionization of $C^0$.

The net effect of these opacity sources is to remove nearly all light shortward of 1060Å, and much of the light around Lyα, as shown in the lower curve in Figure 9b. Thus line of sight clouds *do not* produce an additional strong absorption feature at the Lyman limit. This is because nearly all light shortward of Lyα has already been removed by O I$^*$ absorption. The lack of an absorption feature at the Lyman limit was the reason that clouds were thought to not lie along the line of sight.

To quantify this, in the top panel of Figure 9, we show the net observed spectrum for a representative covering factor of 40%. The red line in Figure 9(a) presents the standard model, with no intervening clouds, while the intervening cloud model is the black line. Intervening clouds produce strong extinction between the Lyman limit and Lyα. These features distinguish between cases with and without clouds along the sight line. These features begin to be obvious when the hydrogen column density reaches ~$10^{23}$ cm$^{-2}$, as shown in Figure 7. Due to the obvious difference of the two models, we believe that this strong absorption can be used to determine whether there are clouds partially blocking the line of sight.

These figures show spectra of a static ensemble of clouds and so have sharp absorption features. In reality the clouds will also have a distribution of velocities, and the absorption features will reflect this distribution. We show simulated spectra with this velocity dispersion next.

Osterbrock & Pogge (1985) find that the Hβ FWHM is 2400 km/s for Mrk 766. We adopt this in our spectral simulations (Figure 10). The red line is the standard model while the black line is the intervening cloud model with 40% covering factor. The blue line is the HST STIS spectrum retrieved from their web site (observation ads/Sa.HST#O5L502030). A redshift of z=1.026 (Rines et al. 2003) was assumed. The HST data were smoothed by averaging over 5Å intervals and negative fluxes were not plotted

The comparison in Figure 10 suggests that intervening clouds do exist in Mrk 766, as R11 found for the X-ray absorber. The HST data have a slope similar to the IBC model. This also suggests that the intrinsic SED of Mrk 766 is more like the red line in Figure 10 and appears softer due to the intervening clouds.

Intervening clouds may exist in other AGNs, where the line widths are broader than Mrk 766. We plot spectra with a FWHM 10000 km/s, which is the typical velocity dispersion of BLR clouds (Peterson 1997), in Figure 11. The red, black, and blue lines represent the standard model spectrum with no intervening clouds, and spectra with intervening clouds having covering factors of 20% and 40% respectively. We plot this range of covering factors because it may change from object to object.

The observational consequences of a range of covering factor can be judged from Figure 11. The different LOS covering factors do not change the emission features, which depend on the global covering factor. The standard model is the same as the IBC model with a 0% covering factor and has the harder SED. The effect of changing the covering factor is to change the SED of FUV continuum. A non-zero covering factor depresses the spectrum around and below 1200Å, with the effect increasing with increasing covering factor.

The feature we predict may also have been seen in surveys. The composite spectrum obtained from the Large Bright Quasar Survey (Francis et al 1991) may suggest that LOS clouds are universal in AGNs. In that Figure 2 of that paper, it is clear that the SED is depressed below Ly$\alpha$, which is very similar to our results for a large LOS covering factor. However, this composite spectrum is strongly affected by the Ly$\alpha$ forest, which is uncorrectable, at wavelengths below Lya.

Observations of low z AGNs (Zheng et al. 2001, Shang et al. 2005) should have far less forest absorption. These show a change in the slope of the SED around the region we predict. Shang et al. (2005) point out that most objects exhibit a spectral break around 1100Å and that this break is intrinsic to AGN. We suggest this break is actually the depression feature predicted by the IBC model. This is qualitatively what we predict: there is a depression feature, in another words, a "break" in the spectrum. Since this break is shown in most AGNs, we suggest that intervening clouds are a common feature in most AGNs. A range in covering factors may explain why the "break" seen by Shang et al. (2005) changes from object to object.

4 CONCLUSIONS

We have simulated the UV/optical and X-ray spectrum produced by an ensemble of BLR clouds that lie along the line of sight to the central engine. We assume the geometry proposed by R11 to explain the observed X-ray variability. Our goal is to quantify what we would see in the optical/UV when BLR clouds lie along the LOS to the central continuum source. We find the following:

1) The UV/optical continuum source has a size much larger than a single cloud. This distinguished the UV / optical spectral region from the X-ray, where a single cloud can cover a large fraction of the emission source. One single cloud crossing the LOS will have almost no observed effect on the UV/optical spectrum since it covers a tiny fraction of the visible accretion disk.
2) It is far more likely that an ensemble of clouds will be present along the sight line, if the emission line regions are to have the deduced global covering factor. In this case there will not be time variability, since we see the ensemble of clouds, but the clouds will produce a net absorption in the optical/UV continuum. The clouds also produce strong emission lines, of course.
3) We predict that a strong broad extinction feature will be present between Lyα and the Lyman limit. This is produced by a combination of Lyα scattering and O I$^*$ absorption in the clouds.
4) There is no spectral break at the Lyman limit because the clouds have continuous optical depths greater than unity for much of the region below Lyα. Instead, the FUV spectrum is depressed by a factor related to the line of sight covering factor of the ensemble of clouds.
5) The comparison of HST data and the IBC model spectra for Mrk 766 suggests that intervening clouds do occur and change the SED.
6) The predicted FUV absorption feature may be present in observations of broad-line AGNs. This suggests that line of sight BLR clouds may be present for most AGNs.

Acknowledgements
Ye Wang thanks IHEP AGN Group for the hospitality of Beijing International AGN summer school and C. Gaskell for his help. We acknowledge support by NSF (0908877; 1108928; & 1109061), NASA (07-ATFP07-0124, 10-ATP10-0053, and 10-ADAP10-0073), JPL (RSA No 1430426), and STScI (HST-AR-12125.01 and HST-GO-12309).

We thank the anonymous referee for their help in improving the manuscript.

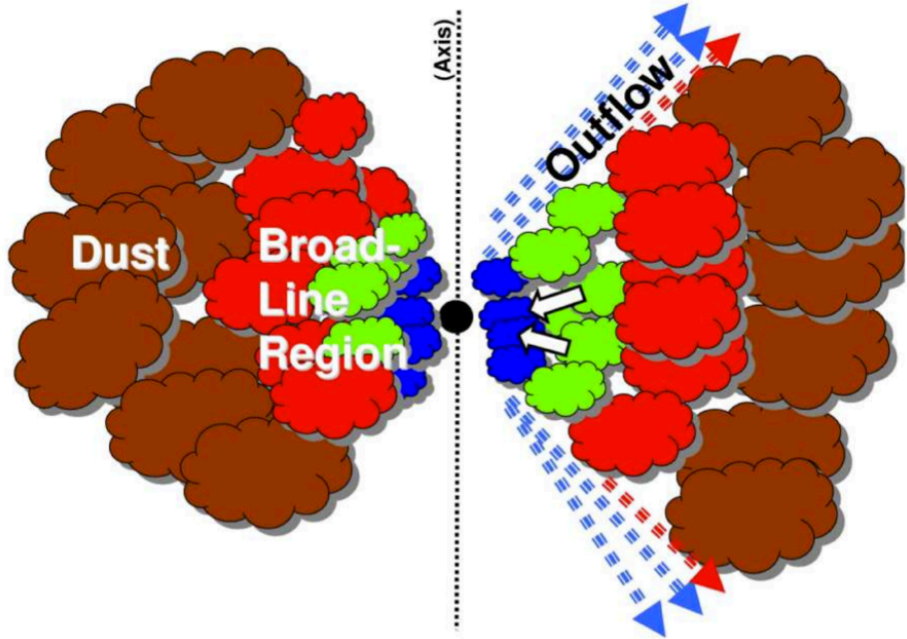

Fig 1. The structure of the broad line regions of AGN (reproduced from Fig 6 of Gaskell (2009) with permission).

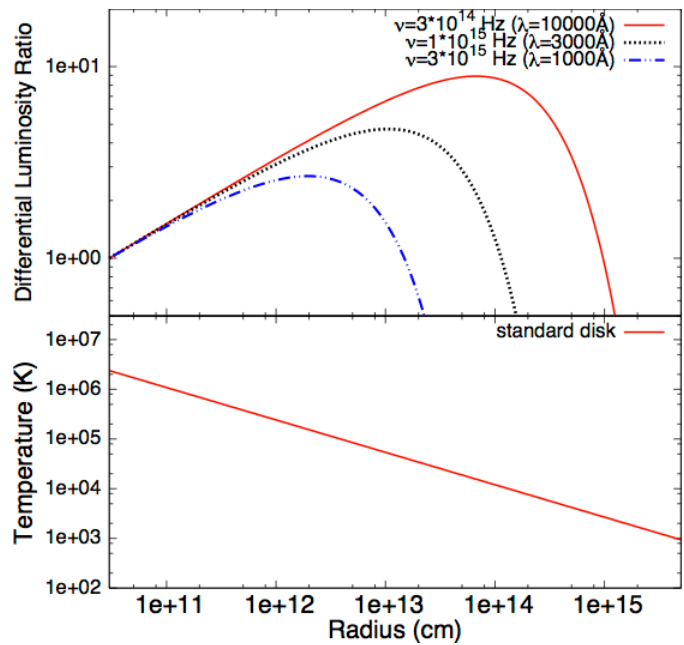

Fig 2. The upper graph shows the differential luminosity ration $\eta_v(r)$ (defined in Sec 2.1) and the lower graph shows the temperature T as the function of radius r for the standard α-disk model described in the text. Parameters of Mrk 766 are used in this figure.

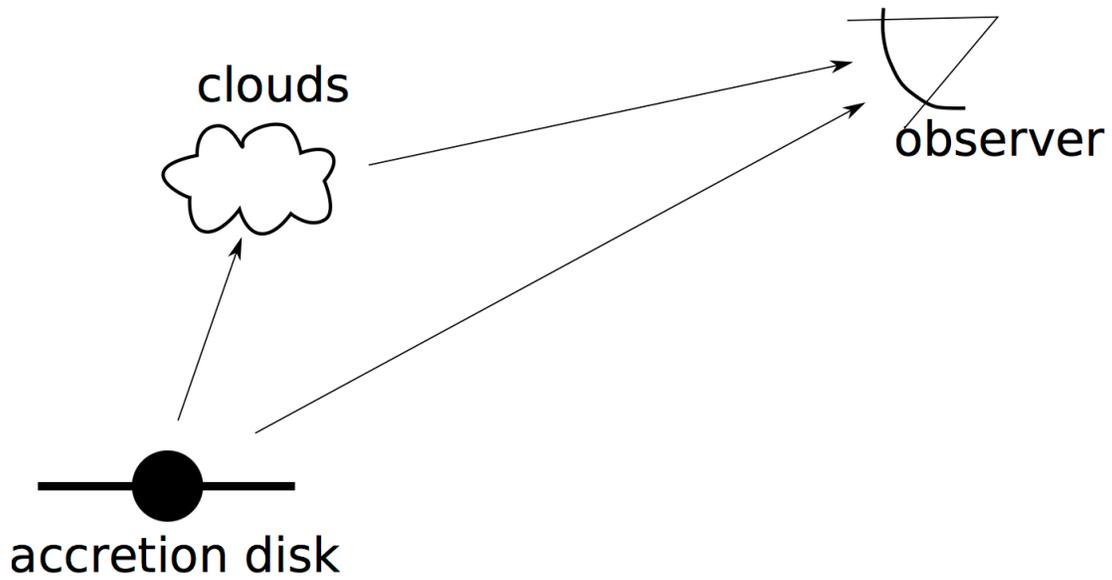

Fig 3. Structure of the standard model. Clouds are not in the line of sight between observer and light source.

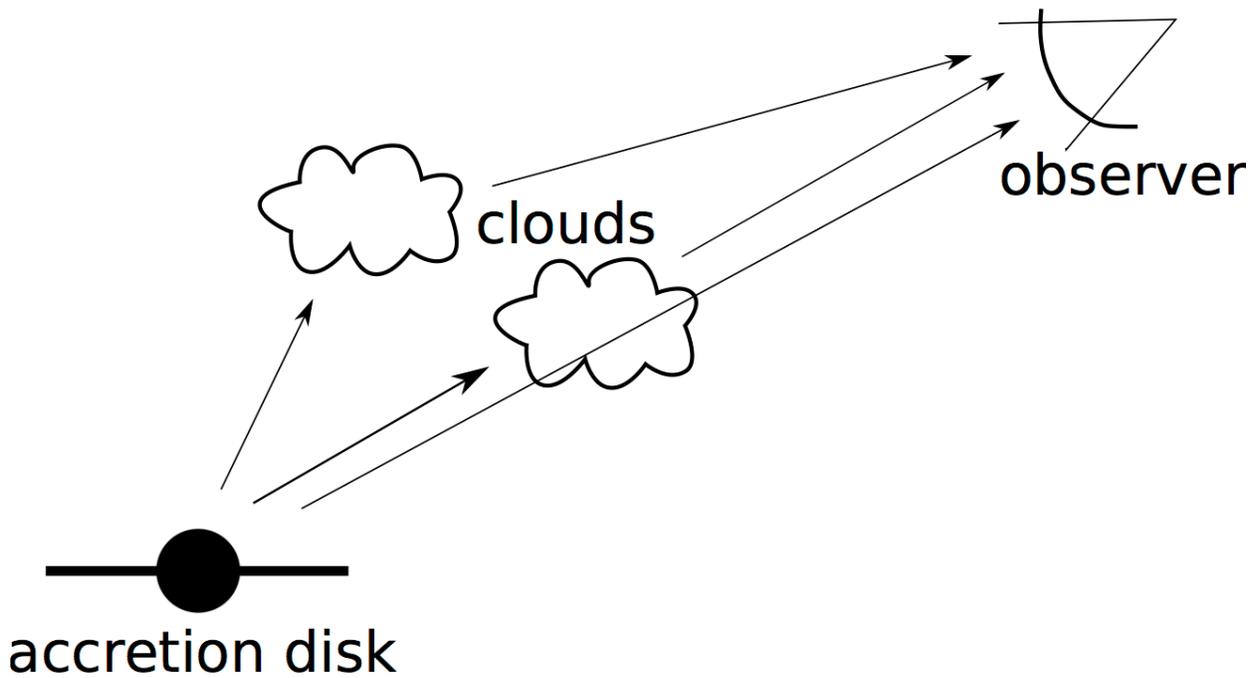

Fig 4. Structure of IBC model. Some clouds lie along in the line of sight between observer and light source.

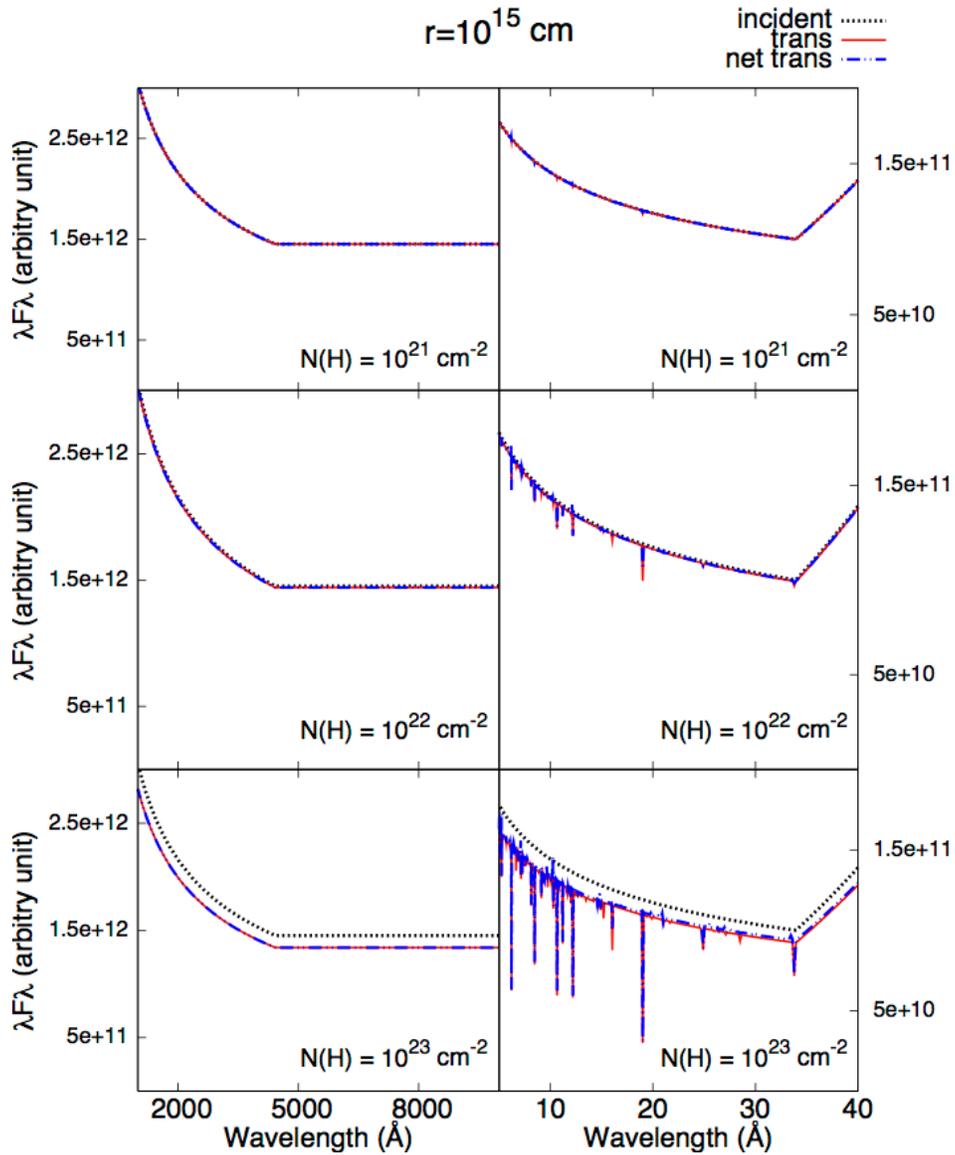

Fig 5. Simulated X-ray (right side) and UV/optical (left side) spectra for clouds at r=$10^{15}$ cm away from the continuum source and column density of cloud is $10^{21}$ cm$^{-2}$, $10^{22}$ cm$^{-2}$ and $10^{23}$ cm$^{-2}$, respectively. The black line is the incident spectrum and would be observed if not be covered by clouds. The red line indicates the spectrum absorbed by clouds but with no emission. The blue line is the transmitted plus emitted spectrum.

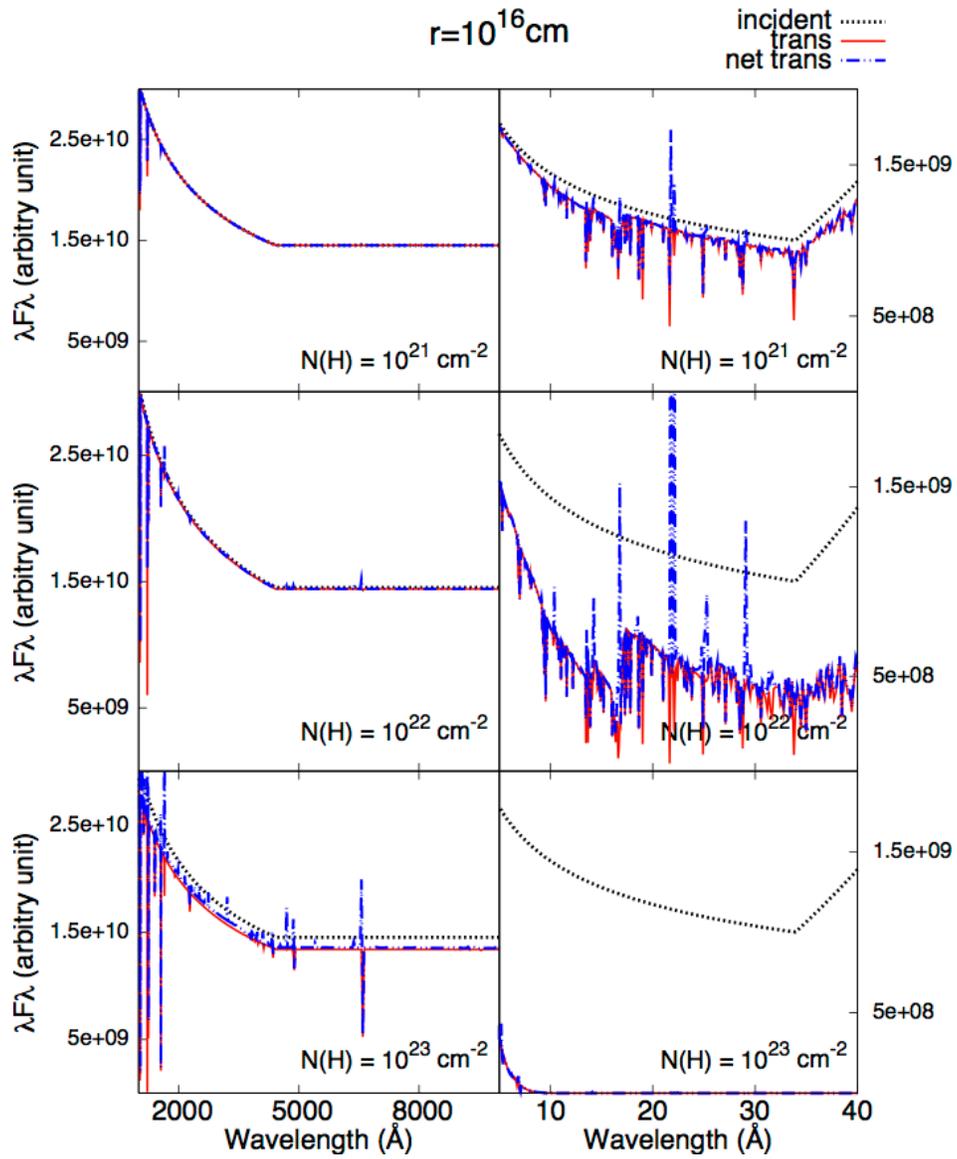

Fig 6. Simulated spectra of X-ray and UV/optical with clouds at r=$10^{16}$ cm away from the continuum source. Others are as same as Figure 5.

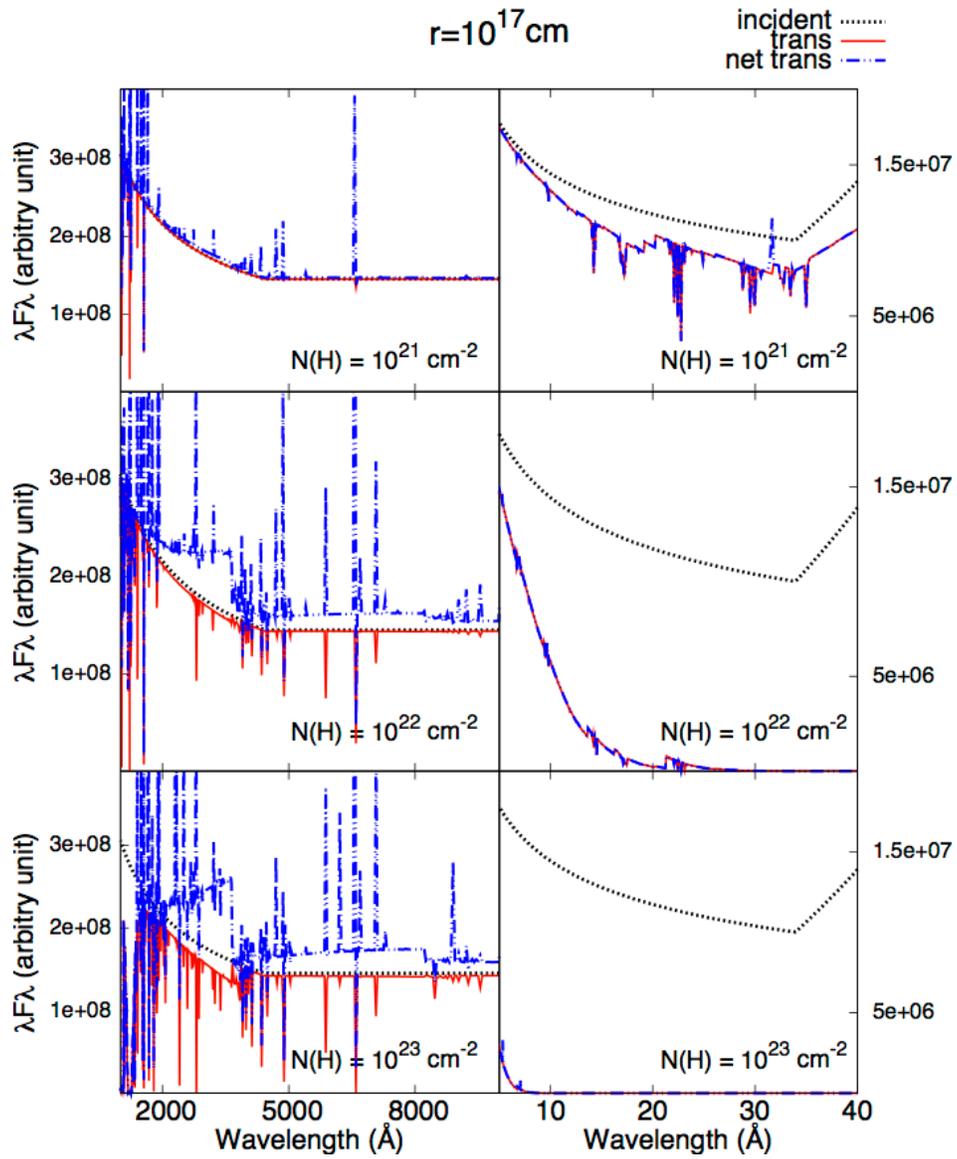

Fig 7. Simulated spectra of X-ray and UV/optical with clouds at r=$10^{17}$ cm away from the continuum source. Others are as same as Figure 5.

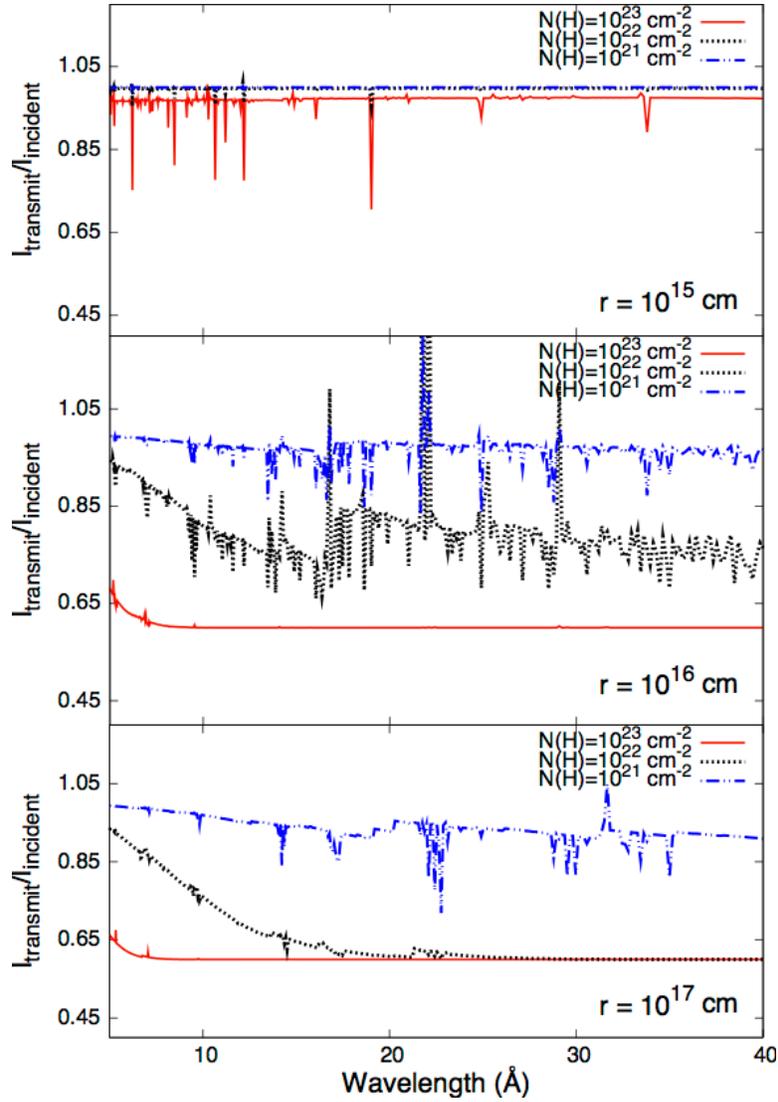

Fig 8. The X-ray transmitted fraction when the cloud covering factor is 40%. The graphs from top to bottom have clouds located at radii r=$10^{15}$ cm, r=$10^{16}$ cm and r=$10^{17}$ cm, respectively. Regions with net emission have transmission greater than unity.

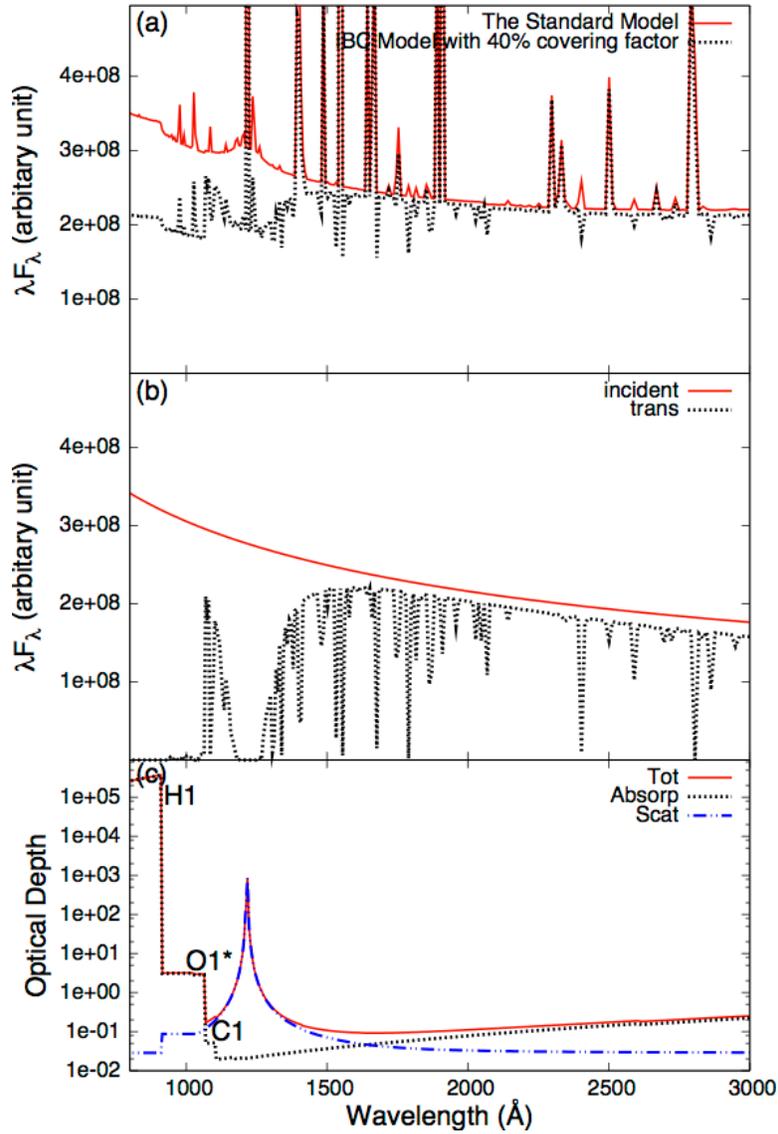

Fig 9. (a) is the simulated spectra for wavelengths between 800Å to 3000Å, which contains the most distinguishing differences between the two models (red line for the standard model, black line for the IBC model with 40% covering factor). (b) is the incident (red line) and transmitted (black line) spectra. (c) is the optical depth (red line for the total optical depth, black line for absorption optical depth, and blue line for scattering optical depth, respectively) , labels in the graph identify the main opacity sources. Neglect the "broaden" feature in this figure, that is due to resolution of simulation code.

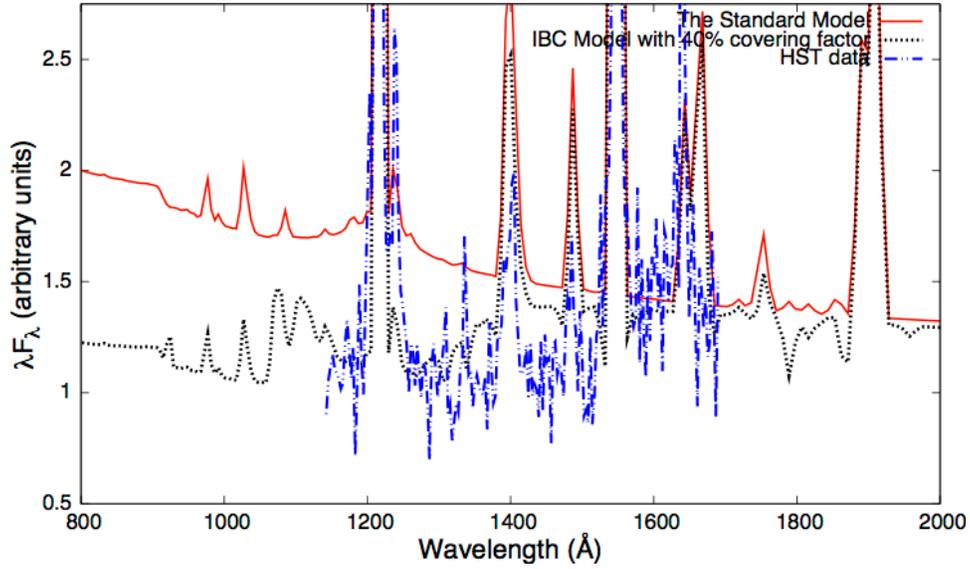

Fig 10. Simulated and observed spectra for Mrk 766. FWHM of simulated spectra is 2400 km/s.

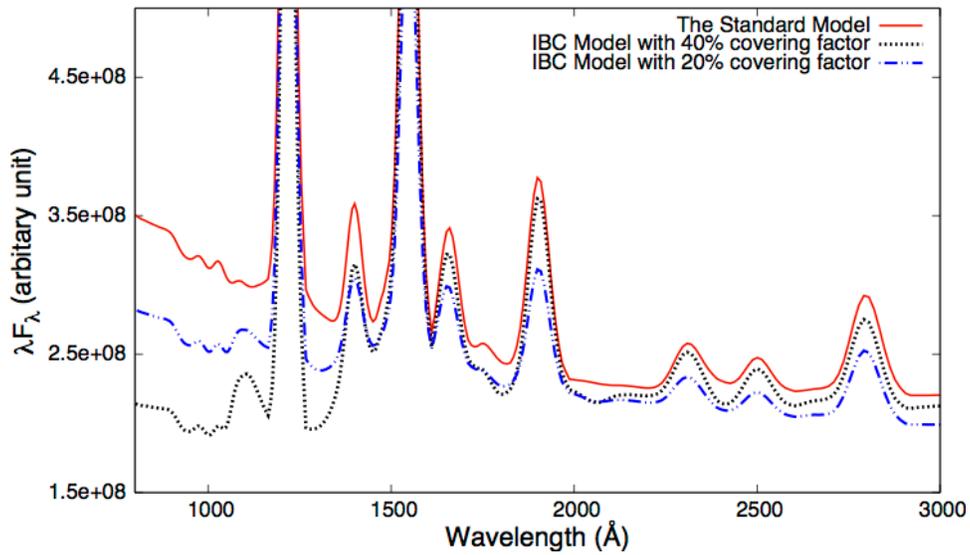

Fig 11. Simulated spectra for broad line AGNs. FWHM is 10000 km/s. The different curves represent covering factors of, from top to bottom, 0, 20%, and 40%.